\documentclass[prl,showpacs,twocolumn,superscriptaddress]{revtex4-1}
\usepackage{amsmath,latexsym,natbib,bm,psfrag,color}
\usepackage[pdftex]{graphicx}
\usepackage{epstopdf}

\begin{document}

\title{Interplay of couplings between antiferrodistortive, 
       ferroelectric, and strain degrees of freedom in 
       monodomain PbTiO$_{3}$/SrTiO$_{3}$ superlattices}

\author{ Pablo Aguado-Puente}

\author{ Pablo Garc\'{\i}a-Fern\'andez}

\author{ Javier Junquera }
\affiliation{ Departamento de Ciencias de la Tierra y
              F\'{\i}sica de la Materia Condensada, Universidad de Cantabria,
              Cantabria Campus Internacional,
              Avenida de los Castros s/n, 39005 Santander, Spain}
\date{\today}

\begin{abstract}
 We report first-principles calculations on the coupling between 
 epitaxial strain, polarization,
 and oxygen octahedra rotations in monodomain
 (PbTiO$_{3}$)$_{n}$/(SrTiO$_{3}$)$_{n}$ superlattices.
 We show how the interplay between (i) the epitaxial strain and (ii)
 the electrostatic conditions, can be used to control the 
 orientation of 
 the main axis of the system.
 The electrostatic constrains at the interface facilitate the rotation
 of the polarization and, as a consequence, we predict large piezoelectric
 responses at epitaxial strains smaller than those that would be
 required considering only strain effects.  
 In addition, ferroelectric (FE) and antiferrodistortive (AFD) modes 
 are strongly coupled. 
 Usual steric arguments cannot explain this coupling 
 and a covalent model is proposed to account for it.
 The energy gain due to the FE-AFD coupling decreases with the periodicity 
 of the superlattice, becoming negligible for $n \ge 3$.
\end{abstract}


\maketitle


 Superlattices composed of thin layers of ferroelectric and paraelectric
 (or incipient ferroelectric) ABO$_{3}$ perovskites oxides have
 attracted a lot of interest during the last few 
 years~\cite{Dawber-05,Ghosez-06,Junquera-08}.
 The fascination for these layered systems comes from
 the fact that the properties of epitaxial structures, made by stacking
 different perovskites, are not a simple combination of the properties
 of the constituent materials, but exotic phenomena might
 emerge that fully rely on interfacial effects.
 
 ABO$_{3}$ perovskite oxides present different phase transition sequences
 and ground states (GS) involving, among others, 
 zone-center ferroelectric (FE) distortions, characterized by the opposite 
 motion of the cations with respect the O cage,
 and non-polar zone-boundary antiferrodistortive (AFD) modes, which consist
 on rotation and tilting of the oxygen octahedra surrounding the 
 B-cation~\cite{Lines-77}.
 But polar and nonpolar instabilities often compete and tend to
 suppress each other, so one of these distortions dominates over the others
 and is the only one that appears in the GS structure. 
 However, the balance is extremely delicate
 and can be tuned by external electrical and strain~\cite{Schlom-07} fields,
 or by changing the chemical environment through the use of different materials
 and periodicities in the stack~\cite{Wu-11}.

 One of the most studied systems in the recent literature, the 
 (PbTiO$_{3}$)$_{m}$/(SrTiO$_{3}$)$_{n}$ (PTO/STO) superlattice,
 constitutes a good example where the balance between different 
 instabilities have been observed to be strongly tunable.
 It was theoretically predicted and experimentally observed that the
 polarization, tetragonality and phase transition temperature of
 the system can be monitored with the number of PTO layers,
 $m$, decreasing monotonically when the PTO volume fraction is 
 reduced~\cite{Dawber-05.2,Dawber-07}.
 However, in the limit of ultrashort periods, PTO/STO  
 superlattices exhibit an unexpected recovery of ferroelectricity 
 that cannot be accounted for by simple electrostatic considerations 
 alone~\cite{Bousquet-08}.
 In this milestone work, the GS of the system 
 was described as not purely ferroelectric, but involving
 a trilinear coupling term between two AFD modes, that correspond to
 in-phase (AFD$_{zi}$) and out-of-phase (AFD$_{zo}$)
 rotations of the oxygen octahedra around the $z$ axis, that induce a
 polar FE distortion (P$_z$) in
 a way compatible with \emph{hybrid improper ferroelectricity}. 

 Despite all the previous efforts, two problems remain
 virtually unexplored.
 First, although many works have dealt independently with the 
 FE-strain and AFD-strain couplings
 (see Ref.~\cite{Schlom-07} and 
 ~\cite{Rondinelli-11}, respectively)
 the influence of the direction and magnitude of
 local polarization on the rotation of the 
 oxygen octahedra has not received the same attention.
 The control over the bond angles through the polarization orientation
 and magnitude could open new routes to generate or enhance magnetoelectric
 couplings~\cite{Benedek-11}.
 Second, first-principles simulations addressing  the transition
 between the improper and normal ferroelectric regimes 
 as a function of the periodicity of the superlattice are,
 up to our knowledge, missing 
 in the literature.
  
 In this Letter we theoretically predict large mixed FE-AFD-strain couplings
 in monodomain PTO/STO superlattices.
 As a result of these interplays, the phase diagram is much 
 richer than originally assumed~\cite{Bousquet-08}, with 
 rotation of the polarization away from the superlattice normal,
 and a strong coupling of the AFD modes with the magnitude and direction 
 of the FE mode.
 We study the physical origin of this phase diagram and the energetics of 
 the FE and AFD contributions and their coupling with increasing periodicity.

\begin{figure}
    \begin{center}
       \includegraphics[width=0.9\columnwidth]{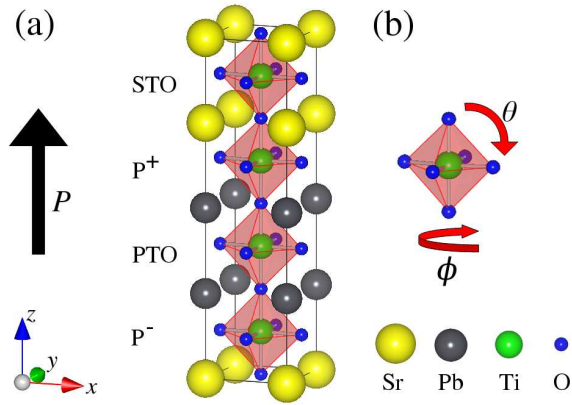}
       \vspace{-10pt}
       \caption{ (color online) 
                 (a) Squematic representation of a (2/2) PTO/STO
                 superlattice. 
                 TiO$_{6}$ octahedra are labeled according 
                 to the chemical identity of the first two neighbor 
                 AO planes, and the direction of the polarization.
                 (b) Definition of the angles of 
                 rotation around the $z$ axis, $\phi$,
                 and tilting around an axis in the $(x,y)$ plane, $\theta$,
                 of octahedra. 
               }
       \label{fig:esquema}
    \end{center}
 \end{figure}

 For this study we perform first principles simulations 
 of (PTO)$_n$/(STO)$_n$
 superlattices,  within the local density approximation (LDA)
 using the {\sc Siesta} code~\cite{Soler-02}.
 Very accurate computations are required since the relevant
 differences in energies are 7 orders of magnitude smaller than
 the absolute value of the energy.
 Real space integrations are
 computed in a uniform grid, with an
 equivalent plane-wave cutoff of 1200 Ry.
 For the Brillouin zone integrations we use a 
 Monkhorst-Pack sampling equivalent to 
 $12 \times 12 \times 12$ in a five atom perovskite unit cell.
 Details on the norm-conserving pseudopotentials and the basis set used
 can be found in Ref.~\cite{Junquera-03.2}.
 The superlattices are simulated by means of
 a supercell approach, where
 we repeat periodically in space 
 a basic unit cell, that is built stacking 
 alternating
 $n$-unit-cells-thick layers of PTO and STO along the [001] direction
 for a global periodicity of ($n/n$)
 [Fig.~\ref{fig:esquema}(a)].
 This structure leads to naturally classify the TiO$_6$ octahedra 
 into four different types 
 (PTO, STO, P$^+$ and P$^-$) depending on the top/bottom AO layer 
 and the direction 
 of polarization [see Fig.~\ref{fig:esquema}(a)].
 In-plane lattice vectors are doubled to 
 account for the condensation of AFD instabilities.
 With the $(2 \times 2)$ in-plane periodicity, TiO$_{6}$ 
 octahedra are allowed both
 to rotate an angle $\phi$ around the $z$-axis or 
 to tilt an angle $\theta$ around an axis contained
 in the $(x,y)$ plane [Fig.~\ref{fig:esquema}(b)].
 The mechanical boundary conditions imposed by the substrate are
 implicitly treated by fixing the in-plane lattice constant, $a_{\parallel}$.
 The use of periodic boundary conditions imposes short-circuit electrical
 condition
 across the whole unit cell.
 
 For every $a_{\parallel}$, a reference non-polar structure is then
 obtained after
 a constrained relaxation 
 within the $P4/mbm$ group,
 until the maximum force
 and $zz$ stress tensor components
 fall below 0.01 eV/{\AA} and 0.0001 eV/\AA$^3$ respectively.
 Then, symmetry is broken displacing coherently the cations by hand,
 and a second relaxation is carried out without any imposed symmetry.

 For the (2/2) superlattice we have performed structural relaxations
 under different in-plane strains, while keeping the global tetragonal
 symmetry.
 The misfit strain is defined as 
 $\varepsilon = \frac{a_{\parallel} - a_{0}}{a_{0}}$,
 where $a_{0}$ is our LDA theoretical lattice constant of cubic
 bulk STO (3.874 {\AA}).

\begin{figure}[htbp]
    \begin{center}
       \includegraphics[clip,width=\columnwidth]{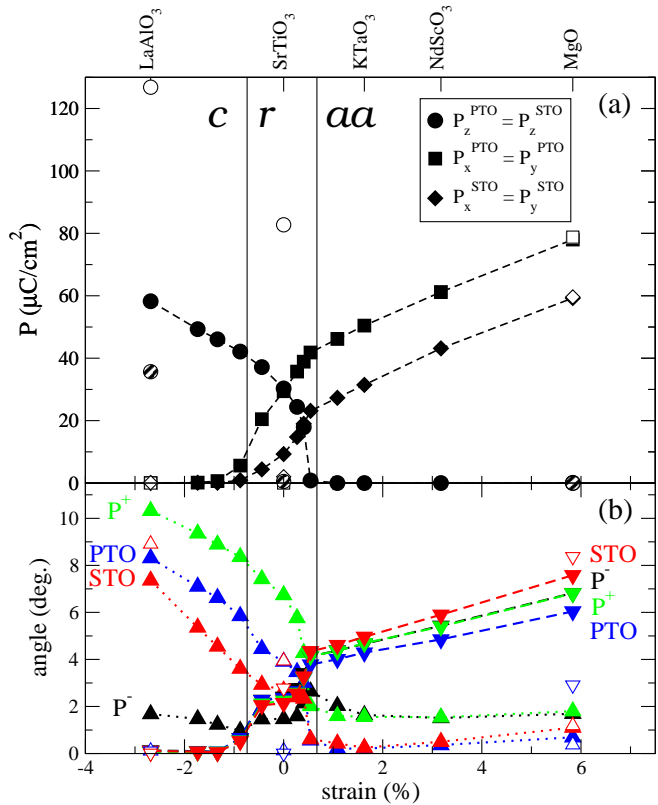}
       \vspace{-20pt}
       \caption{ (color online) (a) Polarization and 
                 (b) absolute value of the oxygen octahedra
                 rotations and tiltings in monodomain
                 (2/2) superlattices under different epitaxial strains.
                 In (a), empty (patterned) circles correspond to bulk 
                 PTO (STO) under the
                 same strain conditions.
                 In (b), the rotation, $\phi$, and tilting, $\theta$, angles
                 of the O octahedra (labeled as in Fig.~\ref{fig:esquema})
                 are represented as triangles-up and 
                 triangles-down, respectively.
                 As in (a), empty symbols refer to bulk values under strain. 
                 Top labels indicate strains induced by common substrates.
               }
       \label{fig:Pvsa}
    \end{center}
 \end{figure}

 The dependence of the polarization 
 (inferred from the bulk Born effective charges and the
 local atomic displacements), and the oxygen octahedra
 rotations and tiltings with the 
 epitaxial strain can be seen in Fig.~\ref{fig:Pvsa}.
 The first conclusion that can be drawn is the existence of a strong
 coupling between FE distortions and AFD modes with strain.
 Both, the magnitude and the relevant directions (of the FE polarization
 and/or the rotation axis of oxygen octahedra)
 can be tuned by the amount of epitaxial strain.

 For large compressive strains the polarization
 along the $z$-direction 
 ($c$-phase in Refs.~\cite{Pertsev-98,Dieguez-04}),
 and the rotations of oxygen octahedra with respect the same $z$-axis 
 are stabilized, while the tiltings are suppressed.
 Besides the effect of the epitaxial strain, the presence of the interfaces
 plays a twofold role, as can be deduced comparing the 
 superlattice and bulk results for pure PTO and STO in Fig.~\ref{fig:Pvsa}:
 (i) the magnitude of the polarization along $z$ is homogeneous throughout
 the heterostructure.
 The polarization mismatch at the interface
 is always smaller than 0.5 $\mu$C/cm$^{2}$,
 highlighting the large electrostatic cost of a polarization
 discontinuity between the layers~\cite{Neaton-03,Dawber-05.2}.
 The price to pay for poling the STO layer is a
 reduction in the polarization of PTO with respect to bulk
 [compare open and filled circles in Fig.~\ref{fig:Pvsa}(a)]. 
 (ii) The magnitude of the angle of rotation is different for the
 different octahedra, a fact that points out a special coupling 
 between the AFD modes and the FE polarization.
 It is remarkable that, although bulk PTO does not exhibit rotation
 of oxygen octahedra, the TiO$_{6}$ octahedra in the superlattice
 with PTO-like 
 environment inherits part of the AFD character of STO and are forced  
 to rotate. 

 In the opposite limit, for large tensile strains, 
 the polarization in the most stable configuration
 lies in the plane, along the [110] direction 
 ($aa$-phase~\cite{Pertsev-98,Dieguez-04}).
 Note that, in this case, there is no electrostatic restriction
 to keep the in-plane polarization at the same value in the
 PTO and STO layers. Therefore, they are decoupled and tend to
 the corresponding bulk values.
 The magnitude of the rotations along the $z$-axis are strongly reduced
 and also tend to bulk, while the tiltings along the [110] 
 axis become the dominant AFD mode. 
 As before, the tiltings in the PTO-like octahedra are enhanced with respect the
 bulk values due to the interfacial coupling with the STO-like octahedra.

 Interestingly, at intermediate strains 
 (around $\varepsilon \approx 0$)
 the polarization
 rotates continuously from the $c$ to the $aa$-phase 
 ($r$-phase~\cite{Pertsev-98,Dieguez-04}).
 Within this regime
 the electromechanical response of the system 
 ($d_{31}$ and $d_{11}$ piezoelectric constants) 
 is enhanced, and at the STO lattice constant amounts to 
 0.30 nC/N, twice larger than the $d_{33}$ in
 Pb(Zr$_{0.5}$Ti$_{0.5}$)O$_{3}$ at room temperature~\cite{Bellaiche-00}.
 The appearance of low symmetry phases where the polarization is
 rotated away from the substrate normal has
 been already reported on PTO thin films grown on DyScO$_{3}$ 
 under tensile strains (+1.4\%)~\cite{Catalan-06}.
 However, we do observe this rotation of polarization and the associated
 enhanced piezoelectric response for much smaller
 strains. This is due to the increased stability of the $r$-phase 
 in this system,
 which is caused by the constrain imposed by the STO
 on the polarization of the PTO layer.
 Our simulations suggest that, 
 in order to reduce $P_{z}$,
 PTO prefers a rotation keeping large the 
 magnitude of $\mathbf{P}$ over a monotonic reduction,
 a well known fact in FE perovskite oxides~\cite{Bellaiche-00,Fu-00}.

 As a summary of the coupling between strain and FE and AFD modes,
 we observe a rotation of the main axis of the system 
 (defined by both the direction of the polarization and the rotation axis of the
 octahedra) 
 from out-of-plane for compressive strains to in-plane for tensile-strains.
 For the AFD modes, this trend can be understood if we consider that the 
 oxygen cage is very rigid. Then, as strains are applied, 
 the system allows the TiO$_6$ octahedra to 
 reorient to maintain their shape
 (see Fig. 3 of Ref.~\cite{Rondinelli-11}).
 More importantly, here we observe an extra \emph{interface coupling} 
 between the FE polarization direction 
 and AFD modes, that reveals itself in the different rotations
 of the octahedra types defined before. 
 The largest difference is observed when strong compressive
 strains are applied: here the P$^+$ octahedra rotate
 more than PTO and STO ones, while P$^-$ octahedra rotate much less.
 In order to understand this result let us discuss the
 origin of the coupling between FE and AFD modes in this system.

\begin{figure}[h]
    \begin{center}
       \includegraphics[clip,width=\columnwidth]{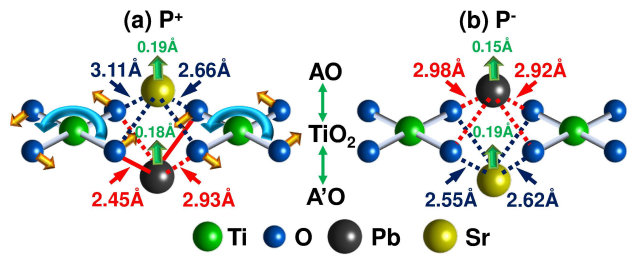}
       \vspace{-20pt}
       \caption{ (color online) 
                 (a) Diagram showing the change in distances between 
                 Pb and Sr ions with O anions at the PTO/STO 
                 interface under compressive strain for 
                 a P$^{+}$ TiO$_{6}$ octahedra.
                 (b) same as in (a) but for a P$^-$ octahedra.               
                 Reduction of distance and reinforcement of the 
                 Pb-O bond is shown by full red lines while an 
                 increase in the Pb-O distance and weakening of 
                 the bond is shown by dotted lines. Blue dotted lines represent
                 Sr-O distances.
                 Green arrows represent out-of-plane displacement of A-cations
                 (Pb or Sr), which magnitude is written in green.
                 Yellow arrows represent the in-plane displacement of O.}
       \label{fig:coupling}
    \end{center}
 \end{figure} 

 AFD distortions are usually regarded as purely steric phenomena, 
 where the rotation of the octahedra takes place if the A-ion is 
 small enough to let the B-O-B bond bend~\cite{Woodward-97}. 
 A polar distortion of the A-cation along the positive $z$-axis
 reduces its distance with the oxygen 
 ions of the TiO$_2$ plane immediately above while increases 
 the distance with the ones below [Fig.~\ref{fig:coupling}(a)]. 
 Taking into account that the ionic radii for Sr$^{2+}$ and Pb$^{2+}$ 
 are almost the same (1.26 and 1.29 \AA, respectively) and using a
 steric model, we would expect that a stronger rotation would only be 
 favored if the mean oxygen-A cation distance increases, leaving space for the 
 octahedra to rotate. However, we find that the mean A-O distance
 is very similar for both P$^+$ and P$^-$ octahedra, 
 close to 2.78\AA\ ,  
 while the rotations are very different.
 In order to explain this difference in rotations, we propose that the 
 driving force in the mixed AFD-FE-strain coupling in PTO/STO superlattices 
 is of covalent nature, instead of steric.
 It is well known that a chemically active
 lone pair on the Pb ion, that allows for strong covalent 
 hybridization with O,
 lies at the origin of FE in bulk PTO.
 For P$^+$, 
 due to the coupling between FE and AFD distortions, not all
 the Pb-O bonds are equivalent.
 In particular, having a shorter (2.45 \AA), and a longer one 
 (2.93 \AA) allows 
 the Pb complex to acquire a pseudo-tetrahedral shape typical of many
 covalent Pb$^{2+}$ compounds.
 For the P$^-$ octahedra, the polar distortion increases the  
 Pb-O distances 
 and the previous mechanism does not apply [Fig.~\ref{fig:coupling}(b)]. 
 These results agree with recent {\it ab-initio} calculations that
 emphasize the role of covalent interactions in the origin of AFD 
 distortions~\cite{GarciaFernandezJPCL10}.
 As the in-plane strain is increased, the polarization rotates away 
 from the $z$ axis and this coupling is reduced making the in-plane 
 rotations small and similar for all octahedra-types when the values 
 of the strain are larger than +1\%. Under these tensile strains,
 the Pb displaces in-plane and both P$^+$ and P$^-$
 become equivalent.

 \begin{table*}[t]
    \caption{ Polarization and relative energies of the different
              monodomain configurations
              for superlattices as a function of
              the periodicity of the supercell.
              In-plane strain corresponds to a STO substrate,
              with a theoretical in-plane lattice constant of 
              $a_{\parallel}$ = 3.874 \AA.
              GS stands for ground state. Polarizations in
              $\mu$C/cm$^2$, energies in meV per five atom perovskite 
              unit cell.}
    \begin{center}
       \begin{tabular}{l|ccc|ccc|ccc}
          \hline \hline
          & \multicolumn{3}{c|}{(1/1)}
          & \multicolumn{3}{c|}{(2/2)}
          & \multicolumn{3}{c}{(3/3)} \\
          & $P_{\rm STO}$ & $P_{\rm PTO}$ & E &
            $P_{\rm STO}$ & $P_{\rm PTO}$ & E &
            $P_{\rm STO}$ & $P_{\rm PTO}$ & E \\
          \hline
          Para.               &
          (0, 0, 0)     & (0, 0, 0)    & +15.3  &
          (0, 0, 0)     & (0, 0, 0)    & +12.8  &
          (0, 0, 0)     & (0, 0, 0)    & +9.7   \\
          $[110]$             &
           (21, 21, 0)  & (31, 31, 0)  & +4.6   &
           (16, 16, 0)  & (35, 35, 0)  & +3.5   &
           (14, 14, 0)  & (36, 36, 0)  & +3.5   \\
          $[001]$             &
           (0, 0, 36)   & (0, 0, 35)   & +1.2    &
           (0, 0, 34)   & (0, 0, 35)   & +3.1    &
           (0, 0, 34)   & (0, 0, 34)   & +4.0    \\
          $[111]$             &
           (14, 14, 32) & (23, 23, 31) & GS      &
           (9, 9, 30)   & (30, 30, 30) & GS      &
           (7, 7, 29)   & (32, 32, 30) & GS      \\
          \hline
          \hline
       \end{tabular}
    \end{center}
    \label{tab:mono_pol}
 \end{table*}

 We have also carried out simulations for different periodicities,
 but fixing $a_{\parallel}$ to the LDA theoretical one of
 STO. 
 We perform both, unconstrained and constrained structural optimizations, 
 where we impose a purely out-of-plane or in-plane polarization
 on the superlattice. Relative energies and polarizations of the 
 PTO and STO layers are gathered in Table \ref{tab:mono_pol},
 while the corresponding rotation angles can be found in the Supplemental 
 materials~\cite{Supplementary}.
 The GS monodomain configuration displays both in-plane 
 and out of plane polarizations, independently of $n$,
 although for $n=1$ the GS $r$-phase reported in Table \ref{tab:mono_pol}
 is essentially degenerated with the $c$-phase (the difference in energy,
 1.2 meV per 5 atom perovskite unit cell, 
 is within the accuracy of our simulations).
 This delicate competition was already observed by
 Bousquet {\it et al.},
 where the phonon frequency of the mode involving in-plane distortions
 in the (1/1) GS of Ref.~\cite{Bousquet-08} 
 was found to be of only 6 cm$^{-1}$, close to become unstable.
 The small difference between the results in Table \ref{tab:mono_pol}
 and those in Ref. ~\cite{Bousquet-08} can be ascribed to small
 changes in the methodology.
 Larger periodicities of the superlattice seem to increase the
 range of stability of the $r$-region, as the difference in energies
 between this phase and the rest increases.
 For $n \ge 2$, within the PTO layer, $\mathbf{P}$ lies close to the
 diagonal of the perovskite unit cell
 (configuration labeled as [111] in Table \ref{tab:mono_pol}).
 The GS can be considered as a condensation of 
 FE$_{z}$ + FE$_{xy}$ + AFD$_{z}$ + AFD$_{xy}$ modes.
 In every case, $P_{z}$ is nicely preserved at the PTO/STO interface,
 with a value between 30 and 35 $\mu$C/cm$^{2}$,
 in good agreement with previous first-principles simulations on 
 clean PbTiO$_{3}$/SrTiO$_{3}$ interfaces~\cite{Cooper-07}.

\begin{figure}[htbp]
    \begin{center}
       \includegraphics[clip,width=0.7\columnwidth]{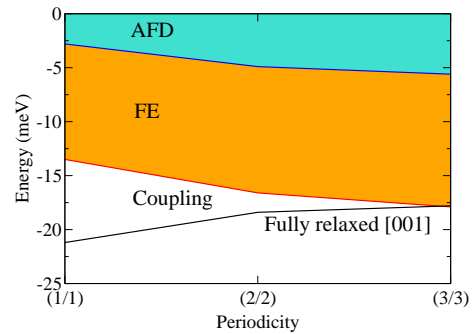}
    \vspace{-10pt}
       \caption{ (color online) Decomposition of the energy
        of the [001] phase into contributions from the polar modes,
        the AFD modes and the coupling between them. The reference
        energy corresponds to a structure where
        neither polar nor AFD instabilities are allowed to condense.
       }
       \label{fig:energy_decomp}
    \end{center}
 \end{figure}

 The structures obtained from the previous relaxations serve also as
 the starting point to answer the question about 
 the evolution of the energy gain due to FE-AFD coupling with the 
 periodicity of the superlattice,
 For this analysis
 we have focused on the [001] phase described above,
 since for this structure the separation of atomic displacements into
 polar distortions and oxygen octahedra rotations is trivial.
 Similar energy decompositions are expected for the GS [111] structure.
 The method to disentangle the different energy contributions 
 due to the polar displacements, oxygen octahedra rotations, and/or their
 coupling 
 can be found in the Supplemental materials~\cite{Supplementary}.
 The results are depicted in Fig.~\ref{fig:energy_decomp} as
 a function of the superlattice periodicity.
 It immediately follows that for ultrashort periodicities 
 the energy contribution from the
 coupling terms is negative and very large, increasing
 significantly the stability of the polarized structure.
 This points in the direction of previous theoretical results~\cite{Bousquet-08}
 that support an ``improper ferroelectricity'' origin for the 
 polarization in these ultrashort periodicities.
 However, the larger the periodicity the smaller the importance of the 
 coupling term, that even changes its sign for 
 $n \geq 3$.
 From this point on, we can consider our superlattices to behave 
 as normal ferroelectrics, in good agreement with experimental results
 [according to Ref.~\citenum{Dawber-05.2}, (3/3) was the threshold 
 periodicity above which the normal FE behavior
 was recovered].

 In summary, our first-principles simulations show how the 
 FE-AFD-strain coupling in monodomain PTO/STO superlattices produces a phase
 diagram much richer than initially envisaged.
 The driving force of the coupling is a combination of 
 electrostatic and covalent effects.
 The new phases might contribute to the stabilization of the monodomain 
 configuration
 over the recently observed and competing polydomain structures~\cite{Zubko-10}.
 The experimental observation of the in-plane component of the polarization 
 in the superlattices remains to be confirmed.

 We acknowledge P. Zubko, Ph. Ghosez, and J.-M. Triscone 
 for the critical reading of the 
 manuscript. 
 JJ is indebted to Dr. J. de la Dehesa for useful discussions.
 Financial support from grants
 FIS2009-12721-C04-02, and 
 CP-FP 228989-2 OxIDes. 
 We acknowledge the computer resources, 
 technical expertise and assistance provided by the 
 Red Espa\~nola de Supercomputaci\'on.
 Calculations were also performed at the ATC group
 of the University of Cantabria.

\end{document}